\documentstyle[12pt]{article}
\begin{document}

\def\pdp{\psi^{\dagger}\psi}
\def\a{\alpha}
\def\b{\beta}
\def\g{\gamma}
\def\d{\delta}
\def\e{\epsilon}
\def\t{\theta}
\def\k{\kappa}
\def\l{\lambda}
\def\om{\omega}
\def\p{\pi}
\def\m{\mu}
\def\n{\nu}
\def\r{\rho}
\def\s{\sigma}
\def\rt{${\tilde r}$}
\def\lnrt{\longrightarrow}
\def\prt{\partial}
\def\prmu{\partial_{\m}}
\def\prlam{\partial_{\l}}

\begin{titlepage}
\hfil{\hfill hep-th/9503144}\break
\vskip .5 true cm
\centerline {\bf{Hyper-Charged Vortices and Strings with
Signature Change Horizon}}
\vskip 1.7 cm
\centerline {\bf{C.S.Aulakh${}^{\dagger}$}}
\centerline{ Department of Physics, Panjab University}
\centerline{ Chandigarh,160014,India}
\vskip 1.1 cm
\centerline{\bf{Abstract}}
\vskip .3 true cm
We show that self-dual Nielsen Olesen (NO) vortices in $3$ dimensions give
 rise to a class of exact solutions when coupled to Einstein Maxwell
Dilaton gravity obeying the Majumdar-Papapetrou(MP) relation between
gravitational and Maxwell couplings , provided certain Chern-Simons
type interactions are present. The metric may be solved for explicitly
in terms of the NO vortex function and becomes degenerate at scales
$r_H \sim l_S exp(\frac{l_S}{l_P})$   where $l_S$ is the vortex core size
and $l_P$ the Planck length. For typical $l_S\geq 10^4 l_P$ the horizon is
thus pushed out to exponentially large scales. In the intermediate
asymptotic region (IAR)  $l_S<<r<<r_H$ there is a logarithmic deviation of
the metric from the flat metric and of the electric field
from that of a point charge (which makes it decrease {\it{slower}} than
$r^{-1}$ :hence the prefix hyper). In the IAR the
ADM energy and charge integrals increase logarithmically with the
distance from the core region and finally diverge at the signature change
horizon. String solutions in $4+p$ dimensions are obtained by replacing the
Maxwell field by an antisymmetric tensor field (of rank $2+p$)
and have essentially similar properties with
$r_H \sim l_S exp((\frac{l_S}{l_P})^{2+p})$
and with the antisymmetric charge playing the role of the topological
electric charge .

\vfil
\vbox{\noindent $\dagger$ On leave of absence from : Institute of Physics,
 Bhubaneshwar and UGC Research Scientistship.\hfil\break
\noindent  {{Email:aulakh@imtech.ernet.in} {\hskip 2.5 cm}\hfil
{\hbox{{March 1995}}}}\hfil\break}

\end{titlepage}
\eject

 In an astonishing recent paper \cite{gibbons} it was shown that the self dual
``instantonic'' solitons of $4+1$ dimensional Yang-Mills (YM) theory and the
self dual BPS \cite{bogo} monopoles of $3+1$ dimensional
Yang-Mills-Higgs theory give rise to a class of exact stable static solutions
even after they are coupled to Einstein-Maxwell(EM)  gravity at the Majumdar-
Papapetrou(MP) point\cite{MP} (i.e the Maxwell coupling is fixed by the
gravitational
coupling so that the static Newton and Couloumb potentials are equal
in magnitude).
Important
additional requirements are the inclusion of certain non-minimal terms of
form $\e^{MNLPQ} A_M F_{NL}^A F_{PQ}^A$ in $4+1$ dimensions (and  its
dimensional reduction in $3+1$ dimensions) and coupling to a dilaton in $3+1$
dimensions. Here  $A_M,A_M^A$ are the Electromagnetic and YM gauge potentials
and $F_{MN}^A$ is the YM field strength . These terms are responsible for
the soliton acquiring a topological electric charge coupled to the Maxwell
field by a mechanism similar to one proposed for $4+1$ Yang Mills
Chern Simons(YMCS) theory solitons in \cite{syn,cts,tsm}.
This  electric charge acts as a Bogomol'nyi bound under the ADM \cite{adm}
mass of field configurations in the theory.
 The self dual solitonic solutions of
Ref.[1] saturate this bound in such a way as to imply the existence of a
Killing spinor with respect to a certain Einstein Maxwell covariant
derivative. Since the bosonic action used is a subset of the $d=5, N=2$
supergravity \cite{gunyadin} it follows that the solutions preserve one
 of the two supersymmetries thus providing yet another example of the
supersymmetry associated with self dual solutions of field equations
\cite{sdsusy}. The result of Ref.[1] allowed us to immedeately confirm
\cite{tep} the  conjecture \cite{cts,tsm} concerning the gravitational
stabilization of instantonic configurations in $SU(N), N\geq 3$ YMCS theory.
 The elegance and naturalness of the arguments of Ref.[1] lead one to expect
 that generalizations to other dimensions and types of solitons (strings,
vortices etc.) should exist. In this letter we obtain solutions
in $3$ and $d=4+p (p\geq 0)$ dimensions analogous  to those
  of Ref.[1] in four and five dimensions. However the peculiarities of
the spherically symmetric
Greens function in two spatial dimensions (i.e the logarithm)
lead to some peculiar properties for the solutions we generate.
The metric decreases logarithmically with the distance from the
 vortex/string core and finally becomes {\it{zero}} at
$r=r_H \sim l_S exp((\frac{l_S}{l_P})^{d-2})$
 ($l_S,l_P$ are the vortex core size and
the Planck length respectively). Strictly speaking, there is thus no
region asymptotic to Minkowski space even though the vortex is
localized on a scale $l_S$. However, in the intermediate asymptotic
regime/region(IAR) $l_S<<r<<r_H$ the metric has only a logarithmic deviation
from flatness with coefficient $O((l_P/l_S)^{d-2})$. Similarly the ADM energy
integral
and  the
charge integral increase slowly with the size of the region of integration $R$
before finally diverging as $R\rightarrow r_H$.
The Majumdar-Papapetrou electric field (in 4 dimensions the antisymmetric
field strength) initially decreases marginally {\it{slower}} than that of
a charged vortex/string (hence the prefix hyper in the title)
 and finally diverges at the signature changing
horizon.

The action we shall treat is the sum of three pieces $S_{gr}, S_{SD}$
 and $S_{CS}$ :

\begin{equation}
S_{gr}= -(16 \p G)^{-1} \int d^3 x E ( R +{e^{2 b \s }} F^2 +
2 (\prt \s )^2)
\end{equation}
\begin{equation}
S_{SD} =  -\int d^3 x{}E{} (\frac{1}{4 g^2}{e^{2 b_a \s }} f^2 +
{e^{2 b_{\psi} \s }} \mid D\psi\mid^2 +\frac{ g^2}{2} {e^{2 b_u \s }}
(\psi^{\dagger}\psi -v^2)^2)
\end{equation}
\begin{equation}
S_{CS} =\int d^3 x\quad \epsilon^{\m \n \l} A_{\m}
({\k}_{1} {\prt}_{\n} a_{\l} +{\k}_2 {\prt}_{\n} J_{\l})
\end{equation}

 $F_{\m\n},f_{\m\n}$ are field strengths of the MP ($A_{\m}$) and NO ($a_{\m}$)
Abelian gauge potentials, $\psi$
the charged scalar field of the NO model, $\s$ the dilaton field and $J_{\l}
={\frac{1}{2i}}\psi^{\dagger}{\stackrel{\leftrightarrow}{D_{\l}}}\psi,
( D_{\l}=\partial_{\l}\psi -i a_{\l}\psi)$ the NO current.
Notice that the MP coupling has been equated to the gravitational one :
$g_e^2 =4 \p G$ The dilaton couplings $b,b_a,b_{\psi},b_u$
will be chosen in the course of the calculation to allow an exact solution
of the full theory given a flat space self dual NO vortex solution.
The other possible Chern-Simons terms
can be added to $S_{CS}$ without affecting our conclusions .
Our conventions for gravitational quantities are those of \cite{weinberg},
E is the determinant of the {\it{dreibein}} while $\e^{\m\n\l}$ is the
3-d antisymmetric tensor density ($\e^{012}=1$).

To proceed we note that the MP form of the metric in $D+1$ dimensions can
be written in a form equivalent to that of \cite{myers} as ($i=1,2$) :

\begin{equation}
ds^2=-\frac{1}{B^{D-2}} dt^2 + B(x^1,x^2) dx^i dx^i
\end{equation}

While this form of the metric can also be derived from considerations
of the existence of a Killing spinor \cite{gibbons}, one can view this ansatz
simply as dictated by the need to ensure that the off diagonal terms of the
Einstein tensor are automatically zero given that the spatial metric is
conformal to the Euclidean one. In $2+1$ dimensions we therefore take
the metric to be $Diag(-1,B,B)$. We also write $B=e^{2\phi}$.

 The field equations are :
\begin{equation}
{\tilde G}_{\m\n} =G_{\m\n} + 8\p G T_{\m\n}(A,\s)=-8\p
G T^{SD}_{\m\n}(a_{\m},\psi)
\end{equation}
\begin{equation}
{\prt}_{\m} (Ee^{2b\s} F^{\m\n})=-4\p G{\e^{\n\m\l}}\prt_{\m}(\k_1 a_{\l} +
\k_2 J_{\l})
\end{equation}

\begin{eqnarray}
\prt_{\m}(E g^{\m\n}\prt_{\n}\s) - (b/2){e^{2b\s}}E F^2
&=&8\p G E((b_a/4g^2) {e^{2b_a\s}} f^2 +
b_{\psi}{e^{2 b_{\psi} \s }} \mid D\psi\mid^2 \nonumber\\
&+& (g^2/2) b_u
 {e^{2 b_u \s }}  (\psi^{\dagger}\psi -v^2)^2)
\end{eqnarray}

\begin{equation}
\prt_{\m}({e^{2b_a\s}} E f^{\m\n}) =-2 g^2 E {e^{2b_{\psi}\s}} J^{\n} -
g^2 \e^{\n\m\l}(\prt_{\m}A_{\l} (\k_1 -\k_2 \psi^{\dagger}\psi) )
\end{equation}
\begin{equation}
D_{\m}({e^{2b_{\psi}\s}} E D^{\m}\psi) ={e^{2b_u\s}} E \frac{\prt U}
{\prt{\psi^{\dagger}}} +
i\k_2 \e^{\n\m\l}\prt_{\n}A_{\m} D_{\l}\psi
\end{equation}
 where $U=(g^2/2)(\psi^{\dagger}\psi -v^2)^2$ is the potential.

When the metric is flat and $\s=0,\k_i=0$ the model reduces to the selfdual
Abelian Higgs  model whose field equations are solved and the energy
minimized provided the fields are static, $a_0=0$,
 and the Bogomol'nyi equations:
\begin{equation}
(D_i \mp i\e_{ij} D_j)\psi =0
\end{equation}
\begin{equation}
 f_{ij} =\pm\e_{ij} g^2 (\psi^{\dagger}\psi -v^2)
\end{equation}
 are satisfied. These first order equations may be decoupled to
yield the well known vortex equation \cite{jaffe} :

\begin{equation}
\prt^2 ln \frac{\psi^{\dagger}\psi}{v^2} =2 g^2 (\psi^{\dagger}\psi -v^2)
+ 4\p \sum_k\mid n_k\mid \d^{(2)}({\vec r}-{\vec r}_k)
\end{equation}
where $\{n_k,{\vec r_k}\}$ are the winding numbers and locations
(positions of the zeros of $\psi$) of an ensemble of (anti)self dual vortices.

We now show that with a certain choice of dilaton weights every solution of
the above self duality equations gives rise to an {\it{exact, explicit}}
 solution of  the field equations of the full theory. We impose the relation :

\begin{equation}
{e^{2b_{\psi}\s}}={e^{2b_a\s}} B^{-1} = B {e^{2b_u\s}}
\end{equation}
which will be satisfied for a certain choice of weights provided $\s=\d \phi$,
$\d$ a constant.
Then it is easy to see, using the assumed flatspace self duality of the NO
fields, that provided
\begin{eqnarray}
A_0&=& \mp\frac{ {e^{2b_{\psi}\s}}}{\k_2 }\nonumber \\
A_i&=&0
\end{eqnarray}
and ${\k_1}/{\k_2}=v^2$
the field equations for the fields $a_{\m},\psi$ are satisfied. Furthermore
it is easy to check that all spatial components of the
 Einstein tensor and the stress tensor of the matter sector vanish . Hence
the spatial components of the Einstein equations are satisfied provided
the spatial stress tensor of the fields $A_{\m},\s$ vanishes  which requires
that
\begin{equation}
\k_2^2=b^2 \qquad\qquad b_{\psi}=-{\frac{b}{2}}
\end{equation}

One finds that three remaining nontrivial field equations
(i.e those for $G_{00}, A_0, \s$) reduce to the {\it{single}} flat space
  equation :
\begin{eqnarray}
\prt^2 e^{2\phi}&=&\pm
(16 \p G)\e_{ij}(\frac{v^2 f_{ij}}{2} +\prt_i J_j)
\nonumber \\
&\equiv&-16\pi G ({\frac{1}{4 g^2}} f_{ij}^2 + \mid D_i \psi\mid ^2 + U(\psi) )
\nonumber \\
&\equiv &(16 \p G) v^2(g^2 (\psi^{\dagger}\psi -v^2) -
\frac{1}{2}\prt^2(\frac{\psi^{\dagger}\psi}{v^2}))
\end{eqnarray}
{\it{if and only if}} one identifies $\s=\phi, b=2$ so that $b_{\psi}=-1,
b_a=0, b_U=-2$. The sign of $\k_2$ can be set without loss of
generality so we put {\hbox{$\k_2=-2$.}}
 Now, using the vortex equation, it immedeately follows that
the regular solution of the above Poisson equation is:

\begin{equation}
e^{2\phi} =C + \m(ln(\frac{\psi^{\dagger}\psi}{v^2})
- \frac{\psi^{\dagger}\psi}{ v^2} -
2\sum_k \mid n_k\mid ln(v^2 \mid\vec r- \vec r_k\mid) )
\end{equation}
where $C$ is an arbitrary constant and $\m=8 \p G v^2 $.
We can fix $C$ to be $1 $ by
noting that as $G\rightarrow 0$ the gravitational,
Maxwell and dilaton fields decouple and therefore we should recover the
Minkowski metric.

In the region outside the core ($r>>\frac{1}{v^2}, \mid r_j\mid$)
$B\rightarrow 1-\m (1+2 N ln(v^2 r));$\hfil\break $ (\sum_k \mid n_k\mid=N)$.
Thus it follows  that if we take G to be positive (see below) then the metric
becomes degenerate at
\begin{equation}
r=r_H= l_S exp({\frac{1}{2 N}} ({\frac{1}{\m}}-1))\approx
l_S exp({\frac{1}{2\m N}})
\end{equation}
where $l_S=1/v^2$ and $l_P=16\pi G$ are the core size and Planck length
respectively. Typically $l_S\geq 10^4 l_P$ (for GUT scale vortices) and can
be as large as $10^{15} l_P$ for weak scale vortices. Thus the signature
changing ``horizon'' at the fantastically large scale $r=r_H$
 does not necessarily make our solutions
(or rather the string solutions in 4 dimensions based on them)
phenomenologically uninteresting since ``frustrated structures '' with
strictly divergent energy (c.f global strings) can  form during phase
 transitions due to the finite size of causally connected domains.

On the other hand , as is well known \cite{jackiw}, the sign of Newton's
constant in 3 dimensions is not fixed {\it{a priori}} since there is no
static gravitational interaction to be made positive . In the present case,
however, $G$ negative would also lead to a ``wrong sign'' for the Maxwell
and dilaton field kinetic terms . The continuation of our solution
 beyond $r=r_H$ requires detailed
analysis (since our solution relies for its consistency on the positivity
of B) and will therefore be taken up elsewhere.

We next consider the  behaviour of our solution
 in the ``intermediate asymptotic region'' ${\frac{1}{v^2}}<<r\leq r_H$ .
The ADM energy  integral over the region
$r\in [0,R]; {\frac{1}{v^2}}<<R\leq r_H$ diverges logarithmically as
$R\rightarrow r_H$ :

\begin{eqnarray}
{\cal{E}}&=&- \int_{r\leq R} d^2 x T^0_0 B = -
\frac{1}{8 \p G} \int d^2 x \prt^2\phi\nonumber \\
&=&-\frac{r B'(r)}{8 G B(r)}{\bigg|^R_0}=\frac{2\pi v^2 N}{1-2\m Nln(v^2 R)}
\end{eqnarray}
where we have assumed that $R$ is much greater than the size of the
region containing the different vortices. Note that the numerator is
precisely the flatspace energy . Since $\m=\frac{l_P}{l_S}$ is so
small we see that for $R<<r_H$ the energy increases very slowly with $R$.
The weak divergence at $r=r_H$ could be cut off by a causal horizon,
as can happen for global strings, or otherwise dynamically smoothed out.

Similarly we can calculate the electric field :
\begin{eqnarray}
F^{0r}&=&\frac{A_0'(r_)}{B}\nonumber \\
&=&\pm\frac{\m N}{r (1-2\m Nln(v^2 r))^3}
\end{eqnarray}

and the charge enclosed in radius $R$ :
\begin{eqnarray}
Q&=&\frac{1}{2}\int_{r=R} F^{\m\n} dS_{\m\n}\nonumber\\
&=&\pm\frac{2 \pi \m N}{ (1-2\m Nln(v^2 R))^2}
\end{eqnarray}

Notice that the enclosed charge is proportional to the winding number as
expected for a topological charge, but it also has the additional peculiar
feature of an $N$ dependence in the denominator.
Clearly both diverge at the signature changing horizon.
Since, by Sylvester's theorem\cite{weinberg}, the number of positive, negative
and zero
eigenvalues of the metric is preserved by a nonsingular general coordinate
transformation it follows that this horizon is not a coordinate artifact.
It would be interesting to continue our solution beyond the horizon and
study the consistency of the propagation of fluctuations on such a background
\cite{townsend}.

String solutions in 4 dimensions are easily obtained from the above analysis
by replacing the Maxwell potential $A_{\m}$ by an antisymmetric gauge field
$B_{\m\n}$ and appropriately modifying the terms involving the antisymmetric
density. Thus one replaces the terms in the action containing $A_{\m}$ by
\begin{eqnarray}
S(B,...)&=&-(16 \p G)^{-1} \int d^4 x E  {e^{2 b \s }} {\frac{1}{6}}
H_{\m\n\l} H^{\m\n\l}\nonumber \\
& +& {\frac{1}{2\sqrt{2}} }
\int d^4 x\quad \epsilon^{\m \s\n \l} B_{\m \s}
({\k}_{1} {\prt}_{\n} a_{\l} +{\k}_2 {\prt}_{\n} J_{\l})
\end{eqnarray}

where $H_{\m\n\l}= \prt_{[\m}B_{\n\l]}$. Now if
i) all fields are independent of  $x^0,x^3$, ii) the only nonzero component of
$B_{\m\n}$ is $B_{03}={\sqrt{2}} A_0$ and the metric is
$diag(-1,B,B,1)$ then the new terms reduce to
\vfil
\eject
\begin{eqnarray}
S(B,...)&=&(8 \p G)^{-1} \int d^4 x E  {e^{2 b \s }} (\prt_i A_0)^2
\nonumber \\
&+ &  \int d^4 x\quad \epsilon^{ij} A_0
({\k}_{1} {\prt}_{i} a_{j} +{\k}_2 {\prt}_{i} J_{j})
\end{eqnarray}

which is exactly the form of the action under our ansatz in the 3
dimensional case apart from the trivial extra integration which is
balanced by the different dimensions of couplings and fields :
$[B_{\m\n}]=[\k_2]=[g]=0,[a_i]=[\psi]=1$ etc. All our results for
vortices thus carry over {\it{mutatis mutandis}} to the 4 dimensional case
with the axionic charge replacing the electric one. Clearly by again raising
the rank of the tensor potential we can embed our basic solution in any
number of dimensions.
Note that in d dimensions $\m={\frac{l_P^{d-2}}{l_S^{d-2}}}$ and so
$r_H\approx v^{-{\frac{2}{d-2}}} e^{\frac{1}{2\m N}}$ grows even
faster with ${1/l_P}$.
The pathologies of 3 dimensions and the special couplings
of our `dilaton' are  responsible for the peculiarities of our solution.
Nevertheless as an addition to the library of exact solutions of
gravitationally coupled theories it merits further investigation.
It motivates
the search for further examples of the mechanism of Ref.[1] which involves
as an essential part the aquisition of {\it{topological electric fields}} by
magnetic configurations in the presence of CS type couplings
\cite{syn,cts,tsm,gibbons}. Moreover our solution may prove of use in
studying the problem of the propagation of fluctuations on spacetimes
with a signature change \cite{townsend}.
We have also found flatspace versions of these results (and
of those for d=4+1 in Ref.[10]) by choosing suitable dilaton couplings.

\vskip 1 true cm
\noindent{{\bf{Acknowledgements}}}
\vskip .5 true cm
I am grateful to Avinash Khare and Pijush Ghosh for useful and
stimulating discussions including one in which they described unpublished
work in which they attempted to obtain charged string solutions in 4
dimensions using $B\wedge F $ couplings of the type used in this paper.

\vskip 1 true cm

\eject
\end{document}